\documentclass[showpacs,preprintnumbers,amsmath,amssymb,nofootinbib]{revtex4}

\usepackage{graphicx}

\usepackage{bm}
\newcommand{\vc}[1]{\bm{#1}}

\newcommand{\ieps}[1]{\includegraphics[height=3in]{#1}}

\newcommand{\cov}[0]{{\rm cov}\;}

\newcommand{\re}[0]{{\rm Re}\;}

\newcommand{\im}[0]{{\rm Im}\;}

\begin{document}

\title{Noise characterization for LISA}
\author{Julien Sylvestre}
\email{jsylvest@ligo.caltech.edu}
\author{Massimo Tinto}
\email{Massimo.Tinto@jpl.nasa.gov}
\affiliation{Jet Propulsion Laboratory, California Institute of Technology, Pasadena, California 91109}

\date{\today}

\begin{abstract}
  We consider the general problem of estimating the inflight LISA
  noise power spectra and cross-spectra, which are needed for
  detecting and estimating the gravitational wave signals present in the LISA data.  For
  the LISA baseline design and in the long wavelength limit, we bound
  the error on all spectrum estimators that rely on the use of the
  fully symmetric Sagnac combination ($\zeta$).  This procedure avoids
  biases in the estimation that would otherwise be introduced by the
  presence of a strong galactic background in the LISA data. We
  specialize our discussion to the detection and study of the galactic
  white dwarf-white dwarf binary stochastic signal.
\end{abstract}


\maketitle

\section{Introduction} \label{intro}
The LISA mission is a collaboration between ESA and NASA to build and
operate a space-based laser interferometer aimed at detecting and
studying gravitational waves (GW) of astrophysical origin.  The current
schedule places the launch around the year 2011, and the 3-10 years
observation period is expected to produce a number of observations of
galactic binaries \cite{galbin}, of massive black hole mergers and
captures, and possibly of a cosmological stochastic background
\cite{LSR}.  Early studies of the data analysis techniques that will
be required to observe these signals are important since they couple
our present knowledge about the sources LISA will be able to observe
to the science requirements guiding the design of the LISA mission
itself.  It is in this context that we present here a study of the
accuracy that can be expected for the characterization of the noise in
the LISA detector once it has been placed in its heliocentric orbit.
This is needed not only for being able to assess the LISA sensitivity
in a region of the frequency band where the stochastic signal from
coalescing galactic binaries would prevent it, but most importantly
for being able to detect weak gravitational wave signals with
confidence. Theoretical modeling and pre-launch performance
measurements of flight-unit instruments will certainly be performed.
However these might not be sufficient to achieve the level of
accuracy required in the analysis of the data collected during the
inflight operation of LISA.

Specifically, we calculate in the low frequency regime
the optimal approximations for the spectra
of the noise affecting the interferometric combinations sensitive to
gravitational waves, by relying {\it only} on all the possible
cross-spectra between the interferometric combinations that are
sensitive to gravitational radiation and the symmetrized-Sagnac
combination, $\zeta$. The rational behind this is that $\zeta$ couples very weakly to
gravitational radiation while it is affected by the same instrumental
noises as the other combinations \cite{Tinto2000}. 
These approximations give a lower bound on
the error that can be achieved by estimators which
make no prior assumptions about the characteristics of the various
noise sources.  Such estimator could be used for carrying out the
full analysis, or for validating the noise models one could adopt for
implementing more sophisticated and efficient data analysis
techniques.  An outline of the paper is given here.

Section \ref{lbdan} presents a review of the LISA baseline design,
which is then followed by a discussion of the noise characterization
problem in section \ref{ifnc}.  In section \ref{mia} we show that
closure for the solution to the noise characterization problem does
not exist, and that only approximate solutions can be implemented.  
In section \ref{nm} we describe the noise model used in our
calculations, which accounts for possible cross-correlations between
pairs of various noise sources.  Although these terms have been
neglected in the literature, they could contribute
to the overall noise budget of the mission, and for this reason
they are included in our analysis.  We then
present in section \ref{ebotep} our calculation of the lower bound on
the estimated noise spectra in the LISA interferometric combinations
that are sensitive to gravitational radiation.  In section
\ref{discussion} we finally provide our comments on the implications
of these bounds for the astrophysical reach of the LISA mission, and
on possible implementations of noise estimation algorithms that would
approach the error bounds we estimate.

\section{LISA baseline design and notation} \label{lbdan}

\begin{figure}
\begin{center}
\ieps{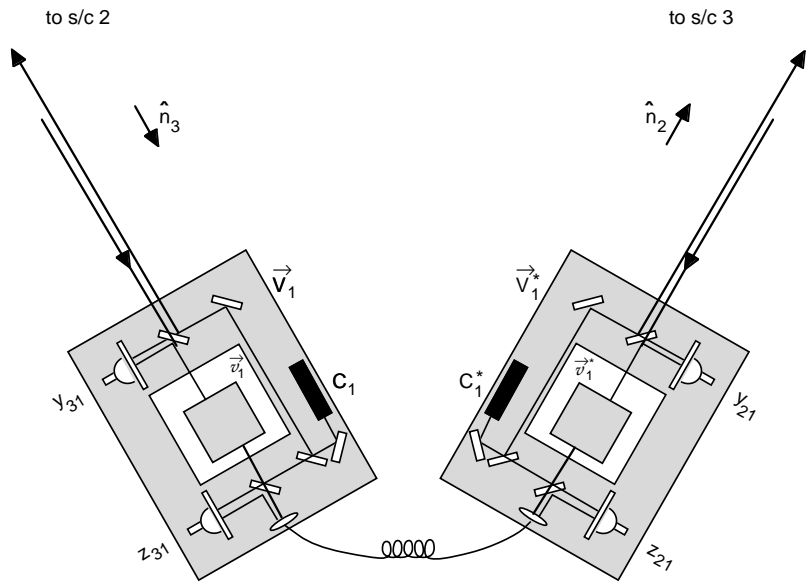}
\end{center}
\caption{Schematic description of the optical benches on spacecraft 1 in the
  LISA baseline design. Variables on the right hand bench are
  distinguished from corresponding variable from the left hand bench
  by asterisks. The measured Doppler signals are $y_{31}$ and
  $y_{21}$. $\hat{\vc{n}}_i$ denote normal vectors along the triangle
  arms, and $\vc{v}_1$ denotes the random velocity vector of proof
  mass 1.}
\label{fig:opticalbenches}
\end{figure}

The LISA mission will consist of three spacecraft flying in an
approximately equilateral triangle formation. The length of the arm
opposite to spacecraft $i$ ($i=1,2,3$) is labeled $L_i$.  The three
nominal arm lengths will be equal to $16.7$ s ($c=1$), and they will
differ from each other by at most a few percents during the entire
duration of the mission.  In contexts where the differences between
the length of the arms can be disregarded, we will simply refer to the
nominal arm length $L$, without indices.  As it is illustrated in
figure \ref{fig:opticalbenches}, each spacecraft will contain two
optical benches.  Each bench will be designed to bounce a laser beam
between each of its proof mass and another proof mass on a distant
spacecraft.  The proof masses are mechanically isolated from their
optical bench, and their relative position is measured with
electrostatic sensors.  These generate inputs for the drag-free
control systems that allow the spacecraft to isolate the proof masses
from external disturbances in such a way to maintain them (to a very
good approximation) in a free-falling configuration in the directions
to the other two spacecraft.  In addition, laser beams will be bounced
off the back of the proof masses and will be exchanged between the
optical benches on a given spacecraft in order to sense the motion of
the optical benches relative to the proof masses.

With this baseline design of the LISA spacecraft, the frequency noise
of the lasers will dominate the secondary noise sources (such as
acceleration noise of the proof masses or optical path length
distortions) by approximately seven orders of magnitude in the
frequency band of interest ($10^{-4} - 1$ Hz).  Time delay
interferometry (TDI) has been proposed \cite{TDI,aet99,TDIWP} as a
robust way for suppressing the frequency fluctuations from the lasers
and the mechanical vibrations of the optical benches to levels smaller
than that identified by the secondary noise sources.
This is done by properly time-shifting and
linearly combining the measurements of the phase differences between
the coherent laser beams exchanged between the three pairs of
spacecraft and the three pairs of on-board optical benches
\cite{TDIWP}.  In reference \cite{aet99} it has been shown that the
entire space of the TDI combinations can be generated by four
combinations, called $\alpha, \beta, \gamma$ and $\zeta$, which are
the Sagnac Interferometric combinations LISA will be able to
synthesize.  Since ($\alpha, \beta, \gamma, \zeta$) can be regarded as
a basis for the entire space of the LISA TDI combinations, in what
follows we will limit our analysis to these four interferometric data.

It should be noted that the modifications of the Sagnac
observable ($\alpha, \beta, \gamma, \zeta$) required in order to account for
the rotation of the LISA array \cite{Shaddock03} and variation in time
of its arm lengths \cite{CH} can be disregarded in the analysis
presented here. Although it will be critical to account for these
effects in order to suppress the laser phase fluctuations below the
secondary noise sources, the modifications introduced to the
gravitational wave and secondary noise responses entering into the TDI
combinations $\alpha, \beta, \gamma$ and $\zeta$ (derived under the
assumption of a stationary LISA configuration) will be negligibly
small.  

It is anticipated that the dominant noises affecting all the TDI
combinations will be introduced by the proof masses and optical path
noises (shot noise and beam pointing fluctuations).  In the notation
of Fig.  \ref{fig:opticalbenches}, the frequency fluctuations generated by the mechanical vibrations of the proof masses are given by \cite{TDIWP} 
\begin{eqnarray}
\zeta^{\rm proof\; mass} = \hat{\vc{n}}_1 \cdot (\vc{v}_{2,2} - \vc{v}_{2,13} + \vc{v}^*_{3,3} - \vc{v}^*_{3,21}) \nonumber \\
+ \hat{\vc{n}}_2 \cdot (\vc{v}_{3,3} - \vc{v}_{3,21} + \vc{v}^*_{1,1} - \vc{v}^*_{1,23}) \nonumber \\
+ \hat{\vc{n}}_3 \cdot (\vc{v}_{1,1} - \vc{v}_{1,23} + \vc{v}^*_{2,2} - \vc{v}^*_{2,13}), \label{eq:anoise}
\end{eqnarray}
where bold characters are either vectors or matrices, $\vc{v}_{i,j} =
\vc{v}_i(t - L_j)$, and $\vc{v}_{i,jk} = \vc{v}_i(t - L_j - L_k)$.
We shall use below the definitions $v_1 = \hat{\vc{n}}_3 \cdot \vc{v}_3$, etc.
Denoting by $n_{ij}$ the optical path noise affecting the Doppler
measurement $y_{ij}$, the optical path noise contribution to $\zeta$
is given by the following expression \cite{TDIWP}
\begin{equation}
\zeta^{\rm optical\; path} = n_{32,2} - n_{23,3} + n_{13,3} - n_{31,1} + n_{21,1} - n_{12,2}. \label{eq:onoise}
\end{equation}
If we denote by $\zeta^{\rm GW}$ the contribution to the $\zeta$
observable from a gravitational wave \cite[Eq. (60)]{TDIWP}, it
follows that $\zeta$ can be written in the following form
\begin{equation}
\zeta = \zeta^{\rm GW} + \zeta^{\rm ins},
\end{equation}
where the instrumental noise part is given instead by
\begin{equation}
\zeta^{\rm ins} = \zeta^{\rm proof\; mass} + \zeta^{\rm optical\; path}.
\end{equation}

In what follows we will be interested in calculating the spectrum of
the $\zeta$ combination
\begin{equation}
S_{\zeta\zeta} = E[\tilde{\zeta} \tilde{\zeta}^*],
\end{equation}
where $\tilde{\zeta}$ is the Fourier transform of $\zeta$, and $E$
denotes the expectation value over many realizations of the
instrumental noise. In the particular case of a stochastic
gravitational wave signal, the expectation value will be performed
over the realizations of the signal too, although it always
will be assumed to be independent of (and therefore uncorrelated with)
the noise. Under these assumptions we can write
\begin{equation}
S_{\zeta\zeta} = S_{\zeta^{\rm GW}\zeta^{\rm GW}} + S_{\zeta^{\rm ins}\zeta^{\rm ins}}.
\end{equation}
In what follows, for any pair of interferometric observables $(x, y)$, $S_{xy}$
will refer to the following function of the Fourier frequency $f$
\begin{equation}
S_{xy} = S^*_{yx} = E[\tilde{x} \tilde{y}^*].
\end{equation}
For simplicity of notation we will write $S^{\rm GW}_{xy}$ for $S_{x^{\rm
    GW}y^{\rm GW}}$ and $S^{\rm ins}_{xy}$ for $S_{x^{\rm ins}y^{\rm
    ins}}$.  
As implied by Eqs.(\ref{eq:anoise},\ref{eq:onoise}), the spectrum $S^{\rm ins}_{\zeta\zeta}$ can be expanded
in terms of proof mass and optical path noises as follows
\begin{eqnarray}
S^{\rm ins}_{\zeta\zeta} = 4 \sin^2 \left(\frac{\omega L}{2}\right) \sum_{i=1}^3 [S_{v_iv_i} + S_{v^*_iv^*_i} + 2 \re S_{v_iv^*_i}] + \sum_{i=1}^2 \sum_{j=i+1}^3 [S_{n_{ij}n_{ij}} + S_{n_{ji}n_{ji}}] \nonumber \\
- 2 \re[S_{n_{12}n_{13}} + S_{n_{21}n_{23}} + S_{n_{31}n_{32}}],
\end{eqnarray}
where $\re x$ denotes the real part of $x$ ($\im x$ is the imaginary part),
and where for the sake of generality we have included terms accounting for
correlations between proof masses on-board the same spacecraft and
between optical path noises along the same arm (see section \ref{nm} for the physical motivations behind the inclusion of these correlations in our analysis).

\section{In-flight noise characterization} \label{ifnc}
As it is well known, the optimal statistics to detect a signal $s$ in
noisy data is the matched or Wiener filter, at least when the noise distribution
can be well approximated as Gaussian. For the combination $\alpha$,
for instance, this statistics is
\begin{equation}
\int df \frac{\tilde{\alpha} \tilde{s}_\alpha^*}{S_{\alpha\alpha}}, \label{eq:matchedone}
\end{equation}
where $s_\alpha$ is the signal $s$ filtered by the transfer function
of gravitational wave to the combination $\alpha$.  As shown in
\cite{pmla02}, the three TDI Sagnac observables $\alpha$, $\beta$, and
$\gamma$ can be combined to form the new data set $A$, $E$, and
$T$. The advantage of these three combinations is in their noises being
statistically uncorrelated for every frequency $f$,
\begin{equation}
E[\tilde{A} \tilde{E}^*] = 0,
\end{equation}
and similarly for all possible pairs taken
among $A$, $E$, and $T$.  The uncorrelated combinations are obtained
by diagonalizing the correlation matrix
\begin{equation}
\vc{C} = \left( 
\begin{array}{ccc}
S_{\alpha\alpha} & S_{\alpha\beta} & S_{\alpha\gamma} \\
S^*_{\alpha\beta} & S_{\beta\beta} & S_{\beta\gamma} \\
S^*_{\alpha\gamma} & S^*_{\beta\gamma} & S_{\gamma\gamma}
\end{array} 
\right),
\end{equation}
leading to an optimal detection statistics for the signal $s$ that is given by
\begin{equation}
\int df \tilde{\vc{x}} \vc{C}^{-1} \tilde{\vc{\chi}}^\dagger, \label{eq:matchedmany}
\end{equation}
where $\vc{x} = [\alpha, \beta, \gamma]$ and $\vc{\chi} = [s_\alpha,
s_\beta, s_\gamma]$,
for $s_\alpha$, $s_\beta$, and $s_\gamma$ the signal $s$ filtered by the
transfer functions of gravitational waves to the combinations $\alpha$,
$\beta$, and $\gamma$ respectively.    
Eq. (\ref{eq:matchedone}) is only a special case
of Eq. (\ref{eq:matchedmany}), with $\vc{C}$, $\vc{x}$ and $\vc{\chi}$
restricted to the one-dimensional case.

The optimal signal-to-noise ratio $\rho_0$ that can be achieved by Eq.
(\ref{eq:matchedmany}) is \cite{pmla02}
\begin{equation}
\rho^2_0 = \int df \tilde{\vc{\chi}} \vc{C}^{-1} \tilde{\vc{\chi}}^\dagger. \label{eq:bestsnr}
\end{equation}
Any variation of the detection statistics (up to a scaling factor) will give a smaller
signal-to-noise ratio, so Eq. (\ref{eq:matchedmany}) is the optimal
detection statistics.  It is also true that the detection statistics
of Eq. (\ref{eq:matchedmany}) maximizes the probability of detecting
the signal for a fixed value of the false alarm probability.

Since the correlation matrix can only be known with some finite
accuracy, we can write it in the following way
\begin{equation}
\vc{C} = \vc{C}_0 + \vc{C}_1, \label{eq:pertC}
\end{equation}
where $\vc{C}_0$ is the true correlation matrix, and $\vc{C}_1$ is the
error in its determination.  By definition, the matrices $\vc{C}$,
$\vc{C}_0$ and $\vc{C}_1$ must all be hermitian and positive definite.
When the correlation matrix is measured with some error, the detection
statistics of Eq. (\ref{eq:matchedmany}) constructed with the measured
matrix is no longer optimal with respect to the true correlation
matrix, and the SNR can be expected to be reduced.  
We present in Appendix \ref{sladipa} a derivation of the SNR loss due to the presence of the error $\vc{C}_1$.
To give an example of the magnitude of this effect, we can
assume the estimate of the correlation matrix to have a given relative
error, $(\vc{C}_1)_{ij} = \epsilon (\vc{C}_0)_{ij}$.
Considering a fairly general signal [cf. Eq.(\ref{eq:wcsnr})], the relative SNR loss is then $\sim \epsilon/2$.
These considerations indicate that in order to limit the SNR reduction to a negligible level (say $\sim 0.1\%$), the relative error on the measured
spectra and cross-spectra for the three TDI combinations $\alpha$,
$\beta$, and $\gamma$ should be of order 10\% at most.

Errors in the determination of the spectra of the TDI combinations
will also have an effect on the measurement of the parameters of the
observed gravitational wave signals.  Let $\vc{I}(\vc{\lambda})$ be the
Fisher information matrix for the estimation of the parameters
$\vc{\lambda}$ using data $\vc{x}$ with distribution function
$p(\vc{x};\vc{\lambda})$:
\begin{equation}
\vc{I}(\vc{\lambda}) = -E\left[ \frac{\partial^2 \log p(\vc{x};\vc{\lambda})}{\partial\vc{\lambda}^2} \right].
\end{equation}
The Cram\'er-Rao bound gives a limit on the error $\cov
\hat{\vc{\lambda}}$ of an unbiased estimator
$\hat{\vc{\lambda}}(\vc{x})$:
\begin{equation}
\cov \hat{\vc{\lambda}} = E[(\hat{\vc{\lambda}}(\vc{x}) - \vc{\lambda})^\dagger (\hat{\vc{\lambda}}(\vc{x}) - \vc{\lambda})] \geq \vc{I}^{-1}(\vc{\lambda}). \label{eq:crbound}
\end{equation}
As usual, the $\geq$ sign in Eq.(\ref{eq:crbound}) means that the left
hand side of the inequality minus its right hand side is positive
semi-definite.  An unbiased estimator can achieve the Cram\'er-Rao
bound if and only if it has the form
\begin{equation}
\hat{\vc{\lambda}}(\vc{x}) = \vc{\lambda} + \left[ \frac{\partial \log p(\vc{x};\vc{\lambda})}{\partial\vc{\lambda}} \right]^T \vc{I}^{-1}(\vc{\lambda}). \label{eq:lambdaest}
\end{equation}
Specializing to the case relevant here, where $\vc{x} = [\alpha,
\beta, \gamma]$ and where $\vc{\chi}(\vc{\lambda}) =
[s_\alpha(\vc{\lambda}), s_\beta(\vc{\lambda}),
s_\gamma(\vc{\lambda})]$, assuming Gaussian noise, and going to the
Fourier domain, one can write Eq.(\ref{eq:lambdaest}) into the following form after some simple algebra
\begin{equation}
\hat{\vc{\lambda}}(\tilde{\vc{x}}) = \vc{\lambda} + {\rm Re}\left[\frac{\partial\tilde{\vc{\chi}}}{\partial\vc{\lambda}} \vc{C}^{-1} (\tilde{\vc{x}} - \tilde{\vc{\chi}})^\dagger \right]^T \vc{I}^{-1}, \label{eq:optestimator}
\end{equation}
where
\begin{equation}
\vc{I} = \frac{\partial\tilde{\vc{\chi}}}{\partial\vc{\lambda}} \vc{C}^{-1} \frac{\partial\tilde{\vc{\chi}}^\dagger}{\partial\vc{\lambda}}. \label{eq:fisherc}
\end{equation}
This estimator achieves the Cram\'er-Rao bound only if the right-hand
side of Eq.(\ref{eq:optestimator}) is independent of $\vc{\lambda}$, as
the notation suggests; if this is not the case, no unbiased estimator
can achieve the Cram\'er-Rao bound.  For the analysis of data from
LISA, one will use the estimator in Eq.(\ref{eq:optestimator}), with
some estimated value of $\vc{C}$ that contains an error, as in
Eq.(\ref{eq:pertC}).  This estimator will necessarily have larger
errors than the Cram\'er-Rao bound.  Because the second term of the
right-hand side of Eq.(\ref{eq:optestimator}) is linear in
$\tilde{\vc{x}} - \tilde{\vc{\chi}}$, $\hat{\vc{\lambda}}(\tilde{\vc{x}})$
will be unbiased, independently of the size of the error $\vc{C}_1$.
It is shown in Appendix \ref{sladipa} that the error term in the covariance of the estimator constructed with the perturbed correlation matrix will vanish to first order in the perturbation $\vc{C}_1$, so that, like the SNR, the variance of the estimated quantities will be at most second order in the measurement errors of the correlation matrix.

At high frequencies ($f \agt 1$ mHz, \cite{tk03}), the gravitational
wave sources are expected to be discrete in time or in frequency, so
that it should be possible to identify portions of the time series
that are free of signal, and can therefore be used to estimate the
spectra with an excellent accuracy (which is limited only by the
integration time, and indirectly by the stationarity of the
noise).  At lower frequencies, on the other hand, gravitational waves
from the galactic binaries will not be distinguishable, and will form
a stochastic signal above the instrumental noise.  Strong signals that
are above this confusion limit might be detected and analyzed with
respect to the noise from the binaries, which can be estimated with
high accuracy by integrating the data for a long enough time.  The
most direct application of the characterization of the noise at low
frequencies is thus the measurement of the stochastic background
itself.  In other words, it is highly desirable to have techniques
that can effectively ``turn off'' the gravitational signal, in order
to properly characterize the instrumental noise.  At higher
frequencies, these techniques should not be crucial for the detection
of weak signals, but should be very useful in understanding and
validating the detector performance. To give a specific example, it
might be useful at some point to verify that a given spectral line is
of astrophysical origin and is not instrumental, without having to
rely on astrophysical assumptions (such as the Doppler modulation of
the signal).

In the simplest search, a stochastic background will be detected by
comparing the measured spectra $S_{\alpha\alpha}$,
$S_{\beta\beta}$, and $S_{\gamma\gamma}$ to the estimate of the
instrumental noise.  Differences between the measured and the
estimated spectra will be interpreted as a stochastic background.  It
therefore goes without saying that the low frequency portion of the
instrumental noise spectra in those TDI combinations must be estimated
as well as possible.

As it has been recognized in \cite{Tinto2000}, the TDI combination
$\zeta$ (i.e., the fully symmetric Sagnac combination) is much less
sensitive to gravitational waves than the other Sagnac combinations
($\alpha, \beta, \gamma$) in the low part of the frequency band
accessible to LISA ($f \alt 30$ mHz), although it is affected by the same instrumental noise
sources.  Consequently, it can be used as a ``gravitational waves
shield'' to estimate the instrumental noise.  This approach was
introduced by \cite{Tinto2000} to bound the power in the stochastic
background. Since the GW signal is additive in the noise of the TDI combinations,
\begin{equation}
S_{XX} = S^{\rm ins}_{XX} + S^{\rm GW}_{XX}, 
\label{eq:insgwsplit}
\end{equation}
where $S_{XX}$ is the measured spectrum of the Unequal-Arm Michelson
combination $X$, $S^{\rm ins}_{XX}$ is the
spectrum of the instrumental noise in $X$, and $S^{\rm GW}_{XX}$ is the
spectrum of a stochastic background. It is shown in \cite{Tinto2000} that the
following inequality holds
\begin{equation}
S^{\rm GW}_{XX} \geq S_{XX} - \hat{S}^{\rm ins}_{XX},
\end{equation}
where the {\it upper bound} on the instrumental noise spectrum is, in the
low-frequency region of the LISA band,
\begin{equation}
\hat{S}^{\rm ins}_{XX} = 16 S_{\zeta\zeta} - 128\pi^2 f^2 L^2 V - 32(3-4\pi^2f^2L^2) N, \label{eq:tintobound}
\end{equation}
with $V$ being a lower bound on the spectrum of the acceleration noise of all
proof masses, and $N$ a lower bound on the spectrum of the optical
noise of all optical paths.

A somewhat similar approach is used in \cite{Hogan2001} to estimate
the {\it average} of three TDI combinations, $\bar{S} = (S_{XX} +
S_{YY} + S_{ZZ})/3$.  They estimate $\bar{S}^{\rm ins}$ by multiplying
$S_{\zeta\zeta}$ by a judiciously chosen transfer function, and
subtract the result from the measured $\bar{S}$ to get $\bar{S}^{\rm
  GW}$.  This ``averaged'' estimator is quite different from what we
will consider below, and will not be discussed further.

\section{Model independent approach} \label{mia}

At low frequencies, the real function $S_{\zeta\zeta}$ and the three
complex functions $S_{\alpha\zeta}$, $S_{\beta\zeta}$, and
$S_{\gamma\zeta}$, give seven measurements that are insensitive to
gravitational waves, in the sense that $S_{\theta\zeta} \simeq S^{\rm
  ins}_{\theta\zeta}$, $\theta = \alpha,\beta,\gamma,\zeta$.  Since
the four TDI generators fulfill the following identity
\begin{eqnarray}
[1-e^{i\omega(L_1+L_2+L_3)}] \tilde{\zeta} = \left[e^{i \omega L_1} - e^{i \omega(L_2 + L_3)}\right] \tilde{\alpha} + \nonumber \\
\left[e^{i \omega L_2} - e^{i \omega(L_1 + L_3)}\right] \tilde{\beta} + \nonumber \\
\left[e^{i \omega L_3} - e^{i \omega(L_1 + L_2)}\right] \tilde{\gamma},
\label{IDENTITY}
\end{eqnarray}
the seven functions introduced above are not all independent. 
In particular, we could try to estimate $S^{\rm ins}_{\alpha\alpha}$ by a linear combination of the cross-spectra $S_{\alpha\zeta}$, $S_{\beta\zeta}$, $S_{\gamma\zeta}$, and $S_{\zeta\zeta}$ as 
\begin{equation}
S^{\rm ins}_{\alpha\alpha} = a S_{\alpha\zeta} + b S_{\beta\zeta} + c S_{\gamma\zeta} + z S_{\zeta\zeta}. \label{eq:saa}
\end{equation}
Using the identity Eq.(\ref{IDENTITY}) to expand the
right-hand side of Eq.(\ref{eq:saa}) and collecting terms in $S^{\rm
  ins}_{\alpha\alpha}$ gives
\begin{equation}
[\cos \omega L_1 - \cos \omega(L_2+L_3)]a + [1-\cos \omega(L_1-L_2-L_3)]z = 1 - \cos \omega(L_1+L_2+L_3). \label{eq:eqn1}
\end{equation}
Further requiring all the other terms in the right-hand side of
Eq.(\ref{eq:saa}) to be zero gives eight other equations.  However,
the only solution satisfying those eight equations is
\begin{eqnarray}
(a,b,c,z) = \left( 
\cos \omega(L_1-L_2)-  \cos \omega L_3, 
1 - \cos \omega(L_1-L_2+L_3), \right. \nonumber \\
\left. -\cos \omega L_1 + \cos \omega(L_2-L_3),
\frac{[\cos \omega(L_1+L_2+L_3) - 1] \sin \frac{\omega}{2}(L_1-L_2+L_3)}{\sin \frac{\omega}{2}(L_1+L_2+L_3)}
\right),
\end{eqnarray}
and this solution does not satisfy Eq.(\ref{eq:eqn1}). This implies
that $S^{\rm ins}_{\alpha\alpha}$ cannot be reconstructed from
measurements of spectra that are insensitive to gravitational waves.
The argument can be extended to other spectra with $\beta$ and
$\gamma$.  It is therefore necessary to consider combinations of the
insensitive spectra that are only approximations to the spectra
involving $\alpha$, $\beta$, and $\gamma$.

\section{Noise model} \label{nm}

In what follows we first define a realistic noise model
for LISA.  The best combination of the spectra and cross-spectra
involving $\zeta$ for this model will be derived in the next section,
where it will be used to set a bound on the accuracy that can be
achieved in characterizing the LISA noise spectra

The principal source of noise at low frequencies is the acceleration
noise of the proof masses along their sensitive axes.  A number of
effects lead to acceleration noises which sum up to a noise budget (per proof mass)
of $3\times 10^{-15}$ m s$^{-2}$ Hz$^{-1/2}$ at 0.1 mHz \cite{PPA}. 
The principal sources of acceleration noise might be due
to \cite{bender}: thermal distortion of the spacecraft, cross-talk
between the other degrees of freedom and the sensitive axis, gravity
gradient noise due to spacecraft displacements, fluctuating spacecraft
magnetic fields, thermal noise, back-action from sensing, electric
forces from fluctuating charge, residual gas, magnetic damping,
temperature variations, etc.  It should be noted that the first four
noise sources might affect {\it both} proof masses on a given
spacecraft, so that the acceleration noises of the two masses will be
correlated to some extent.  This might be important for the correct
estimation of the noise
level that will be achievable with TDI \cite{xcpap}.  In this paper,
we will assume that the acceleration noises, converted in terms of
relative frequency fluctuations, are given by \cite{TDIWP}
\begin{eqnarray}
S_{v_iv_i} = S_{v^*_iv^*_i} = 2.5\times 10^{-48} \left(\frac{1\;{\rm Hz}}{f}\right)^2 \;{\rm Hz}^{-1} \\
S_{v_iv^*_j} = \xi S_{v_iv_i} \delta_{ij}
\end{eqnarray}
where $i,j = 1,2,3$, $\delta_{ij}$ is Kronecker's
delta, and $\xi$ is an arbitrary complex number with norm $|\xi| \leq
1$.  The correlated noises should couple to the proof masses either
gravitationally or magnetically, so the phase angle of $\xi$ will only
depend on the spatial distribution of the perturbing field with
respect to the sensitive axes of the two proof masses.

At frequencies above $\sim 5$ mHz, the optical path noises dominate
the total sensing noise.  The optical path noise budget \cite{PPA}
includes contribution from the shot noise affecting the phase 
measurement at the photodiodes, as well as phase fluctuations resulting from
scattered light effects, laser beam pointing
instabilities, and path length variations on the benches and within
the telescopes.  The latter two effects will couple the optical path
noise along various paths: pointing instabilities will correlate the
noise along the paths related to benches $i$ and $i^*$, while the path
length variations will mostly introduce correlations at times offset
by the one-way-light-time $L$ between the measurements from beams
going in opposite direction along a given arm.  In what follow we will
make the assumption that the latter effect is dominating the cross-correlation
of the optical path noise, and
write the following expressions for the optical path noise spectra and
cross-spectra 
\begin{eqnarray}
S_{n_{ij}n_{ij}} = 1.8\times 10^{-37} \left(\frac{f}{1\;{\rm Hz}}\right)^2 \;{\rm Hz}^{-1} \\
S_{n_{ij}n_{ik}} = -2 \chi \cos(2\pi f L) S_{n_{ij}n_{ij}} \label{eq:conoise}
\end{eqnarray}
where $i,j,k = 1,2,3$, $i \neq j$, $i \neq k$, $j \neq k$, and $\chi$
is a positive real number.  The cosine term in Eq.(\ref{eq:conoise})
comes from the time delay $L$ for the propagation of phase noise along
one of the LISA arms.

In what follows we will use two noise models. In the first case we
will be optimistic and assume all cross-correlation terms to be
negligible, i.e. we will pick $\xi = \chi = 0$.  In the second,
pessimistic case, significant correlations will be assumed to be
present.  For the acceleration noise, we take $|\xi| = 0.25$,
corresponding to the fact that roughly 50\% of the noise budget (in
strain) corresponds to noise sources that can induce strong
correlations between the two proof masses on-board one spacecraft.
Since there is no specific reason for preferring a particular value of
the phase for $\xi$ over any other, we will treat it as a free
parameter in our analysis.  We will also pick $\chi = 0.05$, based on
the fact that roughly 15\% of the optical-path noise budget (in
strain) is attributed to path length variations on the optical benches
and the telescopes.  It should obviously be noted that these noise
models will probably not give a realistic description of the noise
that will affect the LISA measurements.  They are only chosen to
provide some guidance in the noise characterization problem.

\section{Error bound on the estimated spectra} \label{ebotep}

In this section we calculate the smallest possible errors between the
estimated and real instrumental spectra that can be achieved by using
only the GW-free measurements $S_{\theta\zeta}$, where $\theta =
\alpha, \beta, \gamma, \zeta$.  These spectra are the only ones that
are free of laser frequency noise and of GW signals, and therefore
form a natural basis for building noise estimators that do not rely on
unverifiable assumptions about the properties of the noise sources.
Our error bounds will obviously depend on the noise model used, and so
will the estimators that achieve these bounds that we will present.
These estimators will therefore not be valid estimators for the
characterization of the LISA noise; model independent estimators that
approach our error bounds will be presented in a follow-up
communication.

Each TDI combination will show a different transfer function for the
noise contributions identified in the previous section.  In
particular, every spectrum will be a linear combination of the
acceleration and optical noise spectra, so that any instrumental
noise spectrum $S^{\rm ins}_{\mu\nu}$ ($\mu, \nu$ are any TDI combinations) can be written as
\begin{equation}
S^{\rm ins}_{\mu\nu} = \vc{T}_{\mu\nu} \cdot \vc{S}_N,
\end{equation}
where $\vc{T}_{\mu\nu}$ is a complex vector of frequency dependent
coefficients, and $\vc{S}_N$ is the real vector of the $21$ noise
spectra
\begin{eqnarray}
\vc{S}_N = (
S_{v_1v_1}, S_{v^*_1v^*_1}, \re S_{v_1v^*_1}, \im S_{v_1v^*_1}, \nonumber \\
S_{v_2v_2}, S_{v^*_2v^*_2}, \re S_{v_2v^*_2}, \im S_{v_2v^*_2}, \nonumber \\
S_{v_3v_3}, S_{v^*_3v^*_3}, \re S_{v_3v^*_3}, \im S_{v_3v^*_3}, \nonumber \\
S_{n_{12}n_{12}}, S_{n_{13}n_{13}}, S_{n_{12}n_{13}}, \nonumber \\
S_{n_{21}n_{21}}, S_{n_{23}n_{23}}, S_{n_{21}n_{23}}, \nonumber \\
S_{n_{31}n_{31}}, S_{n_{32}n_{32}}, S_{n_{31}n_{32}}
).
\end{eqnarray}
The $S_{\theta\zeta}$ spectra only have terms of the form
$S_{v_iv_i}+\re S_{v_iv^*_i}, S_{v^*_iv^*_i}+\re S_{v_iv^*_i}$, or $\im
S_{v_iv^*_i}$.  This is because $\zeta$ has the same transfer function
to $\vc{v}_{i}$ and to $\vc{v}^*_i$, so any spectrum $S_{\theta\zeta}$
will only have terms of the form $S_{v_iv_i} + S^*_{v_iv^*_i}$ and
$S_{v^*_iv^*_i} + S^*_{v_iv^*_i}$.  Consequently, only three of the
four quantities $S_{v_iv_i}, S_{v^*_iv^*_i}, \re S_{v_iv^*_i}$ and $\im
S_{v_iv^*_i}$ can be
measured simultaneously.  A similar argument applies for the optical
noises.  In that case, a $S_{\theta\zeta}$ spectrum measures the
combinations $S_{n_{ij}n_{ij}} - S^*_{n_{ik}n_{ij}}$ and
$S_{n_{ik}n_{ik}} - S_{n_{ik}n_{ij}}$, so one can only measure
$S_{n_{ij}n_{ij}} - \re S_{n_{ik}n_{ij}}$, $S_{n_{ik}n_{ik}} - \re
S_{n_{ik}n_{ij}}$, and $\im S_{n_{ik}n_{ij}}$ simultaneously.
Inequalities in the lengths of the arms may break these symmetries.

The coefficient vector $\vc{T}_{\mu\zeta}$ is known from the definition
of the TDI combinations, and it just consists of geometrical transfer
functions.  Clearly, there is not enough information to measure every
one of the 21 terms in $\vc{S}_N$.  This means that an
approximated solution has to be used.  Supposing we want to measure
the instrumental spectrum $S^{\rm ins}_{\mu\nu}$, we can measure
$S_{\alpha\zeta}$, $S_{\beta\zeta}$, $S_{\gamma\zeta}$, and
$S_{\zeta\zeta}$, and then approximate $\vc{T}_{\mu\nu}$ with the
following expression
\begin{equation}
\hat{\vc{T}}_{\mu\nu} = a \vc{T}_{\alpha\zeta} + b \vc{T}_{\beta\zeta} + c \vc{T}_{\gamma\zeta} + z \vc{T}_{\zeta\zeta}. \label{eq:Tapprox}
\end{equation}
The estimated spectrum $\hat{S}^{\rm ins}_{\mu\nu}$ is then given by 
\begin{eqnarray}
\hat{S}^{\rm ins}_{\mu\nu} = \hat{\vc{T}}_{\mu\nu} \cdot \vc{S}_N \\
 = a S_{\alpha\zeta} + b S_{\beta\zeta} + c S_{\gamma\zeta} + z S_{\zeta\zeta}.
\end{eqnarray}
Performing the approximation in Eq.(\ref{eq:Tapprox}) requires the
definition of a proper error metric.  In this paper, we choose a model
dependent metric in order to bound the achievable error on the
estimated spectrum.  For complex $\vc{T}_{\mu\nu}$ vectors, we minimize
the error on ${\rm Re}[\vc{T}_{\mu\nu}]$ and ${\rm Im}[\vc{T}_{\mu\nu}]$
independently. Specifically, we minimize the error $\|{\rm
  Re}[\hat{\vc{T}}_{\mu\nu}] - {\rm Re}[\vc{T}_{\mu\nu}]\|^2$ and $\|{\rm
  Im}[\hat{\vc{T}}_{\mu\nu}] - {\rm Im}[\vc{T}_{\mu\nu}]\|^2$, where ${\rm
  Re}[\hat{\vc{T}}_{\mu\nu}]$ and ${\rm Im}[\hat{\vc{T}}_{\mu\nu}]$ are
restricted to the space spanned by the vectors ${\rm
  Re}[\vc{T}_{\mu\zeta}]$ and ${\rm Im}[\vc{T}_{\mu\zeta}]$ respectively,
and where
\begin{equation}
\|\vc{x}\|^2 = (\vc{x},\vc{x}) \label{eq:norm}
\end{equation}
for the scalar product
\begin{equation}
(\vc{x},\vc{y}) = \sum_{i=1}^{21} x_i (\vc{S}_N)_i^2 y_i. \label{eq:dotproduct}
\end{equation}
This error norm penalizes error on individual noise components
proportionally to their contribution to the spectrum $S_{\mu\nu}^{\rm
  ins}$.  The approximation casted in this form is a linear
least-squares estimation problem, which can be solved easily using
standard techniques.  In particular, the vectors
$\vc{T}_{\alpha\zeta}$, $\vc{T}_{\beta\zeta}$, $\vc{T}_{\gamma\zeta}$,
and $\vc{T}_{\zeta\zeta}$ are not linearly independent, and therefore
the solution $\hat{\vc{T}}_{\mu\nu}$ is not unique.  We pick the solution
with the smallest Euclidean norm, as in \cite[Eq. (2.6.7)]{NR}.

\begin{figure}
\begin{center}
\ieps{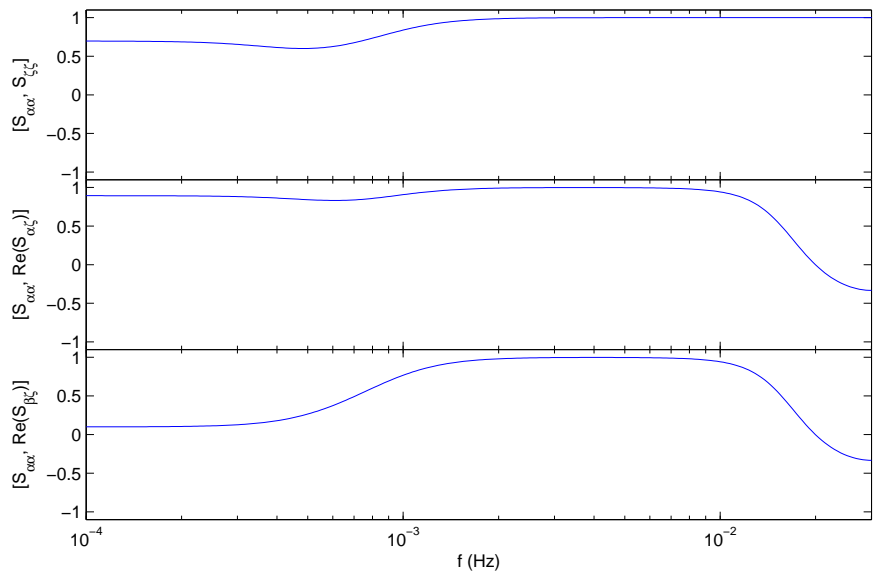}
\end{center}
\caption{Overlaps between the instrumental noise $S^{\rm ins}_{\alpha\alpha}$ and the measured spectra $S_{i\zeta}$. By symmetry, the overlap $[S^{\rm ins}_{\alpha\alpha},{\rm Re}(S_{\gamma\zeta})]$ is identical to the overlap $[S^{\rm ins}_{\alpha\alpha},{\rm Re}(S_{\beta\zeta})]$ (third panel). Overlaps with the imaginary parts of the complex cross-spectra are all zero.}
\label{fig:Cofalphaalpha00overlap}
\end{figure}

\begin{figure}
\begin{center}
\ieps{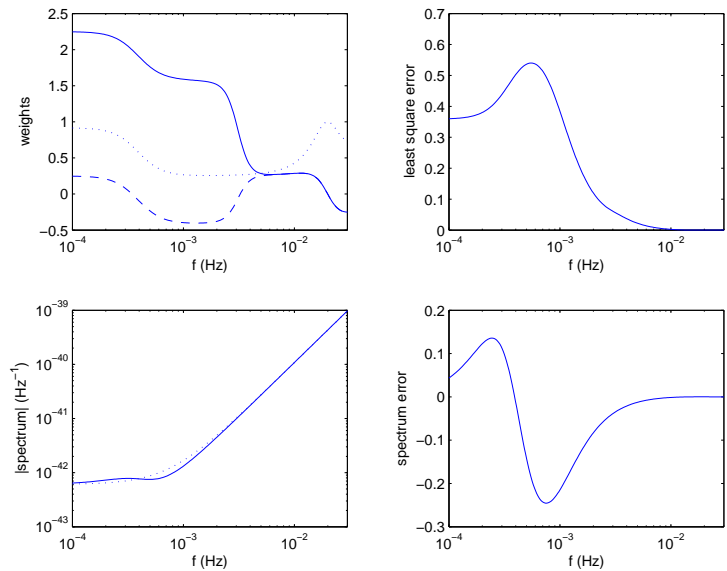}
\end{center}
\caption{Results of the least square fit of $S^{\rm ins}_{\alpha\alpha}$. Top left: the weights $a$ (continuous line), $b$ (dashed line), $c$ (dashed line), and $z$ (dotted line). Top right: the normalized least square error. Bottom left: $S^{\rm ins}_{\alpha\alpha}$ (dotted line) and $\hat{S}^{\rm ins}_{\alpha\alpha}$ (continuous line). Bottom right: the normalized error on the estimated spectrum.}
\label{fig:Cofalphaalpha00}
\end{figure}

\begin{figure}
\begin{center}
\ieps{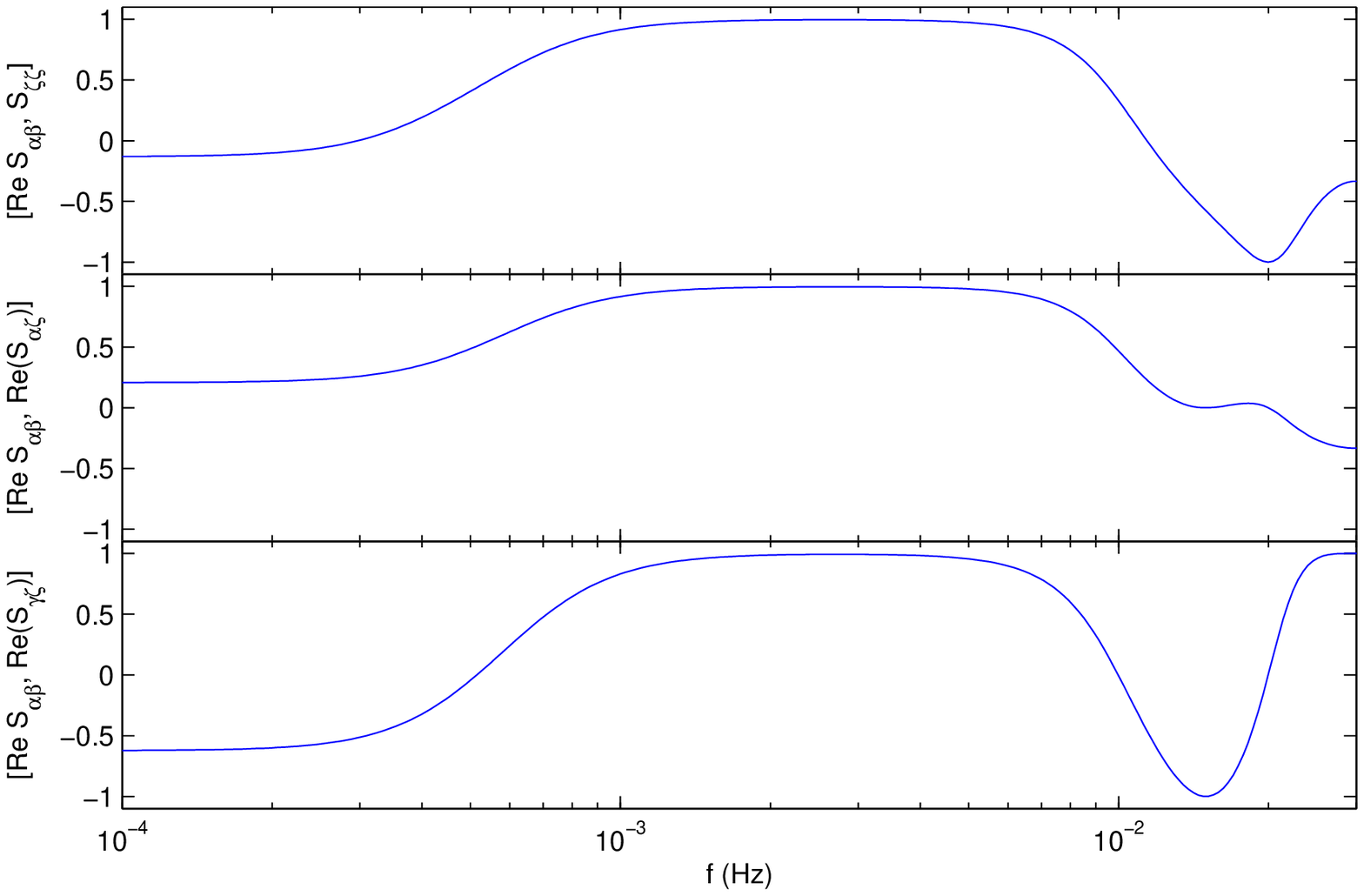}
\end{center}
\caption{Overlaps between the instrumental noise $\re S^{\rm ins}_{\alpha\beta}$ and the measured spectra $S_{i\zeta}$. By symmetry, the overlap $[\re S^{\rm ins}_{\alpha\beta}, \re S_{\beta\zeta}]$ is identical to the overlap $[\re S^{\rm ins}_{\alpha\beta},\re S_{\gamma\zeta}]$ (third panel).}
\label{fig:ReCofalphabeta00overlap}
\end{figure}

\begin{figure}
\begin{center}
\ieps{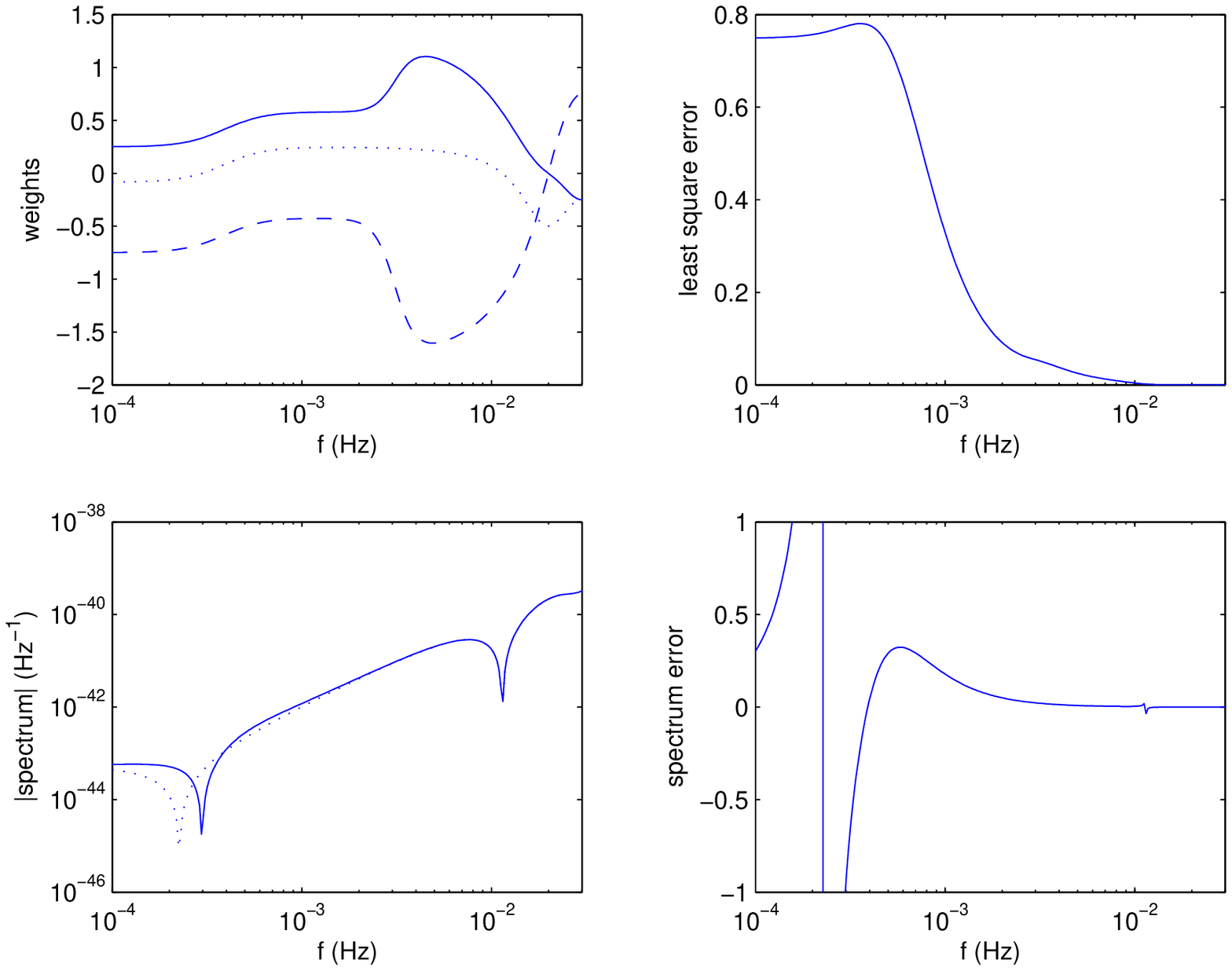}
\end{center}
\caption{Results of the least square fit of $\re S^{\rm ins}_{\alpha\beta}$. Top left: the weights $a$ (continuous line), $b$ (continuous line), $c$ (dashed line), and $z$ (dotted line). Top right: the normalized least square error. Bottom left: $\re S^{\rm ins}_{\alpha\beta}$ (dotted line) and $\re \hat{S}_{\alpha\beta}$ (continuous line). The dips in these curves are unresolved zeros. Bottom right: the normalized error on the estimated spectrum.}
\label{fig:ReCofalphabeta00}
\end{figure}

In order to explore the influence of correlations on the measurement
of the spectra, we consider the two choices of $\xi$ and $\chi$
defined in section \ref{nm}.  For the case of no correlations, we show
in figure \ref{fig:Cofalphaalpha00overlap} the overlaps
\begin{equation} [S^{\rm ins}_{\alpha\alpha}, S_{\theta\zeta}] =
  \frac{(S^{\rm ins}_{\alpha\alpha}, S_{\theta\zeta})}{\sqrt{\|S^{\rm
        ins}_{\alpha\alpha}\|^2\|S_{\theta\zeta}\|^2}}.
\end{equation}
Below $\sim 2$ mHz, none of the $S_{\theta\zeta}$ have very good
overlap with $S^{\rm ins}_{\alpha\alpha}$, but above that frequency,
$S^{\rm ins}_{\alpha\alpha}$ is very close to $S_{\zeta\zeta}$.  We
can thus expect a good least square fit for $f \agt 3$ mHz, which is
what Fig. \ref{fig:Cofalphaalpha00} shows.  The relative least square
error is
\begin{equation}
\frac{\|\hat{\vc{T}}_{\alpha\alpha} - \vc{T}_{\alpha\alpha}\|}{\|\vc{T}_{\alpha\alpha}\|}, \label{eq:lserror}
\end{equation}
and the relative error $e$ on the spectrum is
\begin{equation}
e = \frac{\hat{S}^{\rm ins}_{\alpha\alpha} - S^{\rm ins}_{\alpha\alpha}}{S^{\rm ins}_{\alpha\alpha}}. \label{eq:terror}
\end{equation}
The spectrum error peaks at -24.5\% at 0.76 mHz, although it is small
in some low-frequency bands.
The other non-trivial spectrum for $\xi=\chi=0$ is $\re S^{\rm
  ins}_{\alpha\beta}$.  All cross-spectra are real, and spectra
involving $\beta$ and $\gamma$ are identical to $S^{\rm
  ins}_{\alpha\alpha}$ or $S^{\rm ins}_{\alpha\beta}$.  The overlaps
are shown in Fig.\ref{fig:ReCofalphabeta00overlap}, while the relative
errors are shown in Fig.\ref{fig:ReCofalphabeta00}.  Again, the
overlaps are good for $f \agt 1$ mHz, and the least square and
spectrum errors are small.  The spectrum $S^{\rm ins}_{\alpha\beta}$,
however, has a zero around 0.22 mHz, which cannot be fitted by the
$S_{\theta\zeta}$ measurements.  This leads to the large relative error on
the spectrum below 1 mHz.

\begin{figure}
\begin{center}
\ieps{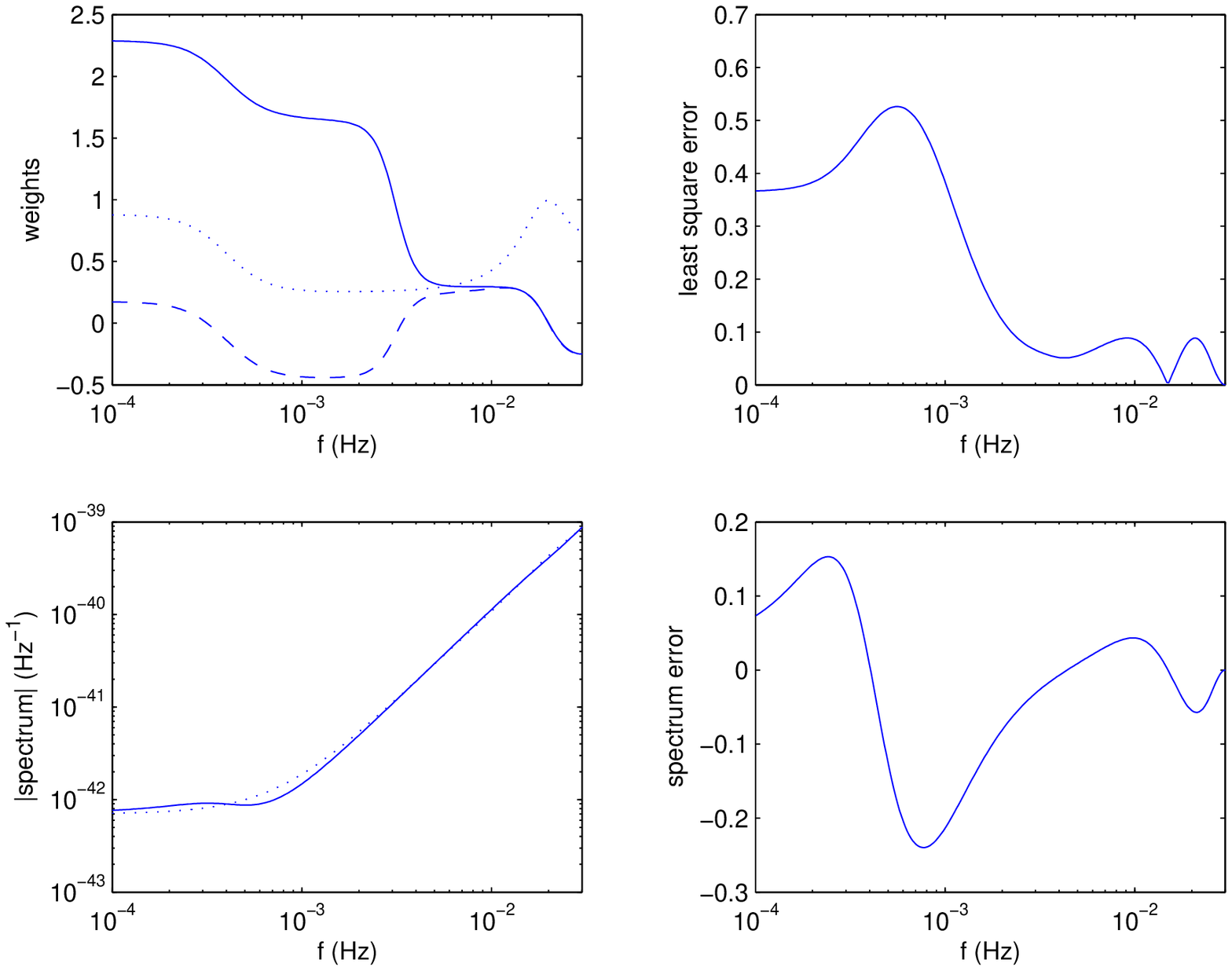}
\end{center}
\caption{Same as Fig.\ref{fig:Cofalphaalpha00} for $\xi = 0.25$ and $\chi = 0.05$.}
\label{fig:ReCofalphaalphaR}
\end{figure}

\begin{figure}
\begin{center}
\ieps{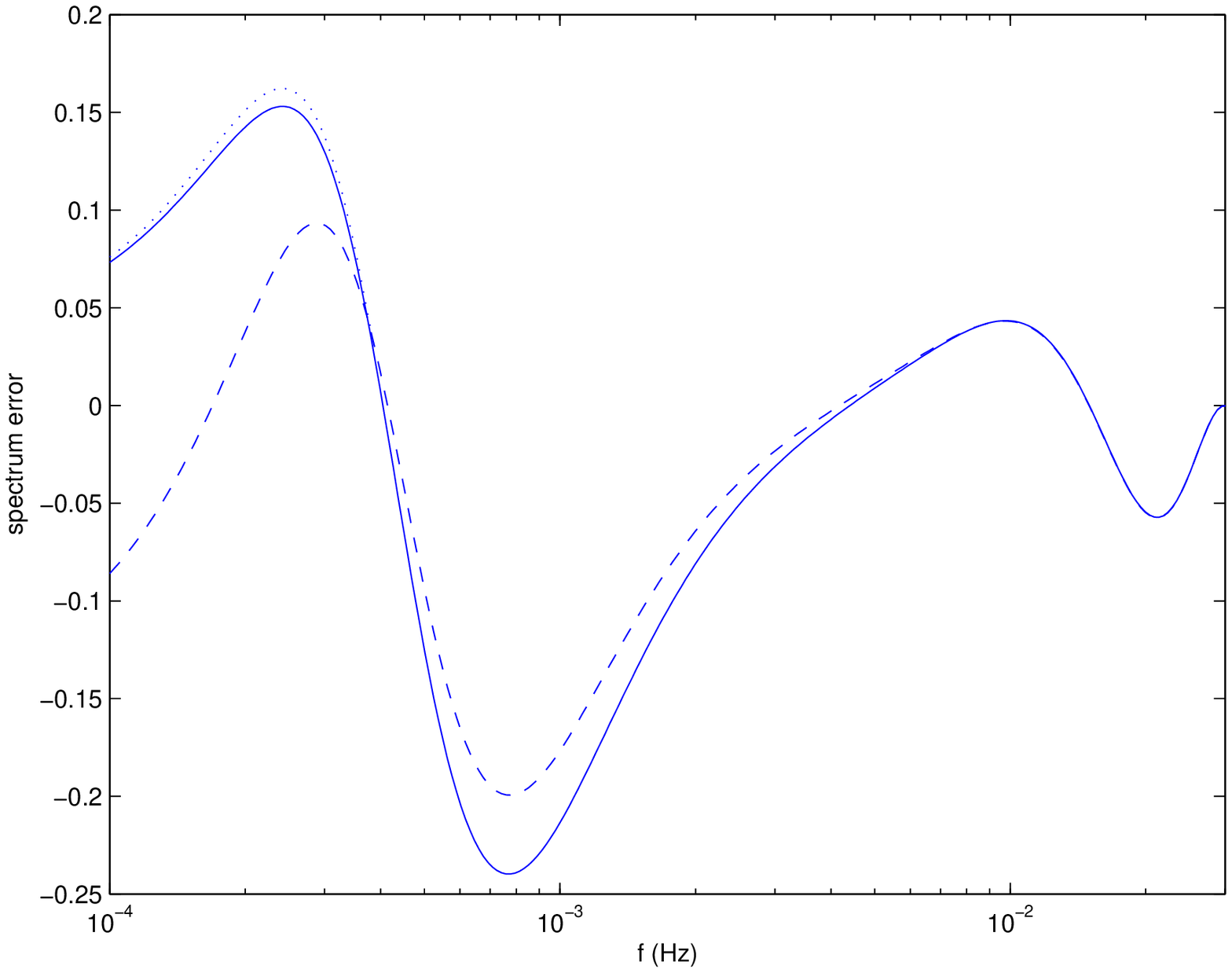}
\end{center}
\caption{Normalized error on the estimated spectrum, for different values of the arguments of $\xi$. The continuous line is for $\arg\xi = 0$, and the dashed line is for $\arg\xi = \pi$. These two curves bound the variation of the error at all frequencies above 0.39 mHz. Below 0.39 mHz, a larger error (dotted line) is possible for a complex $\xi$.}
\label{fig:Cofalphaalphamaxargchixi}
\end{figure}

For the pessimistic case with $|\xi| = 0.25$ and $\chi = 0.05$, the
characteristics of the approximation varies with the arguments of
$\xi$.  As argued in section \ref{nm}, a zero argument for $\chi$ is
probably realistic.  To get a specific example, we also fix the
argument of $\xi$ to be zero, and plot the results of the least square
fit for $S^{\rm ins}_{\alpha\alpha}$ in Fig.
\ref{fig:ReCofalphaalphaR}.  It can be seen that at high frequencies
the error is increased significantly by the small amount of
correlation in the optical path noise.  At low frequencies,
correlations do not appear to significantly increase the error of the
fit.  Fig. \ref{fig:Cofalphaalphamaxargchixi} shows how the error
varies with the phase angles of $\xi$, thus revealing the importance
of the acceleration noise correlations.  Depending on the argument of
$\xi$, the estimation relative error can vary by as much as 10\% at
low frequencies.

\begin{figure}
\begin{center}
\ieps{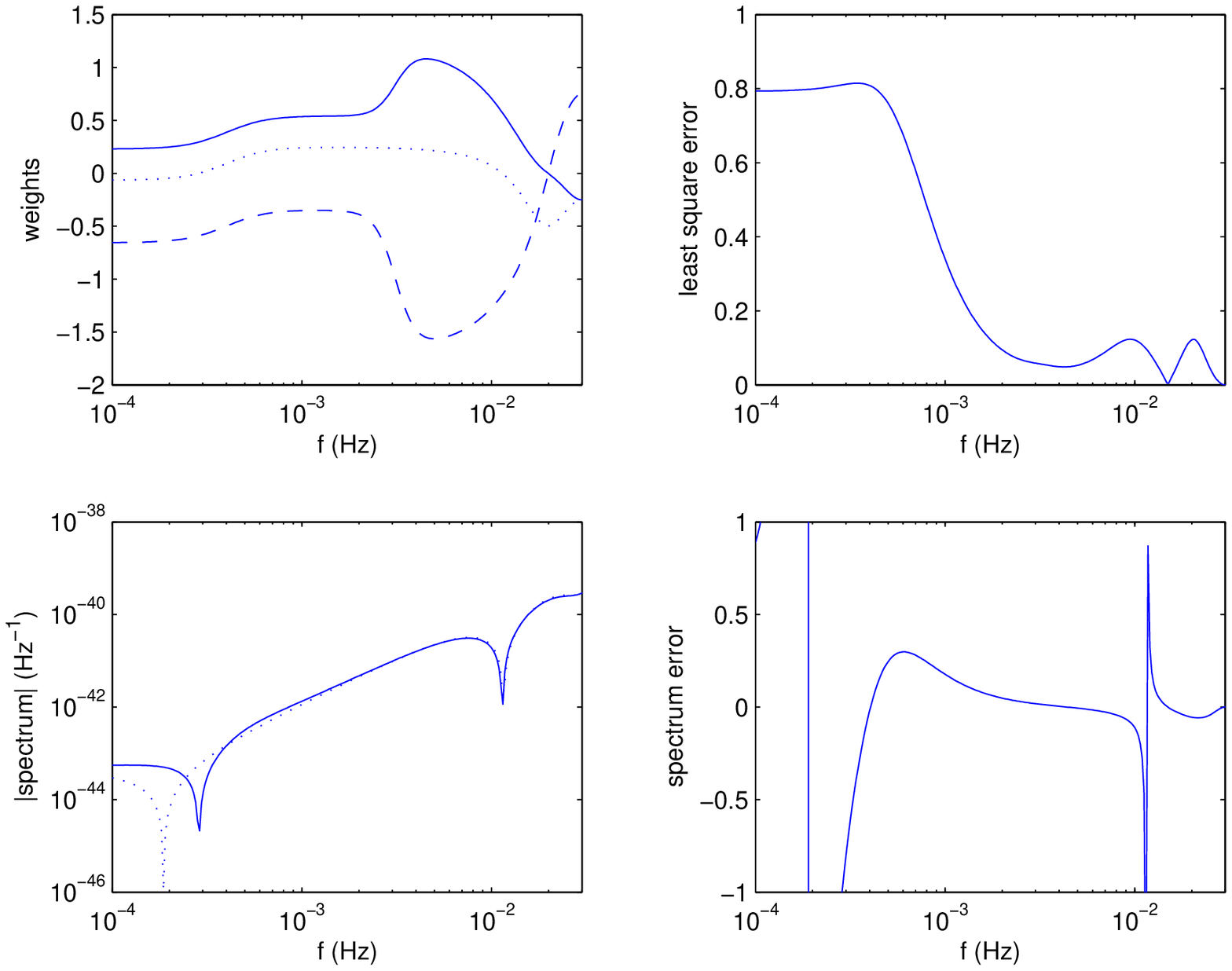}
\end{center}
\caption{Same as Fig. \ref{fig:ReCofalphabeta00} for $\xi = 0.25$ and $\chi = 0.05$.}
\label{fig:ReCofalphabetaR}
\end{figure}

In general, some cross-spectra might be complex, but for the noise
model of section \ref{nm}, they will all be real, if the phase
of $\xi$ is zero.  We present in Fig. \ref{fig:ReCofalphabetaR} the
results of the least square fit for $\re S^{\rm ins}_{\alpha\beta}$,
for the case $\xi = 0.25$ and $\chi = 0.05$.  As it can clearly be
seen, the correlations exacerbate the problem of fitting the zero near
0.22 mHz.

\section{Discussion} \label{discussion}
We have presented an analysis of the prospects for a precise
characterization of the noise of the LISA detector once it begins to
operate as a gravitational wave detector.  We have derived a set of
bounds on the error of the spectra and cross-spectra of the TDI
combinations that will be necessary for the analysis of the data.  The
most important feature of these bounds is that they are derived almost
without any assumptions about the form of the individual noise
sources.  The only exception is that we assume certain
cross-correlations to be negligible in order to keep a manageable
number of terms; there are, however, no fundamental limitations to
including additional terms in future analyses.  The bounds that we
obtained simply reflect how well a given spectrum can be approximated
by a linear combination of spectra that are insensitive to
gravitational waves, and therefore only sample the instrumental
noise.

Our error bounds are different from those given in \cite{Tinto2000} and 
\cite{Hogan2001}, which assumed pre-launch measurements or estimates of
the acceleration and optical path noises to be available in
constructing their estimators of the spectra. In contrast, we have
calculated the optimal error bound that can be achieved by an
estimator that does not use prior assumptions about the spectra of the
various noises.  These bounds will be most useful for the development
of model-independent estimators, which will be the subject of a
follow-up paper.  It is important to emphasize that model-independent
estimators will be critical for the unambiguous detection and study of
GW sources with LISA, and possibly for the validation of the
instrument and of other model-dependent estimators.

\begin{figure}
\begin{center}
\ieps{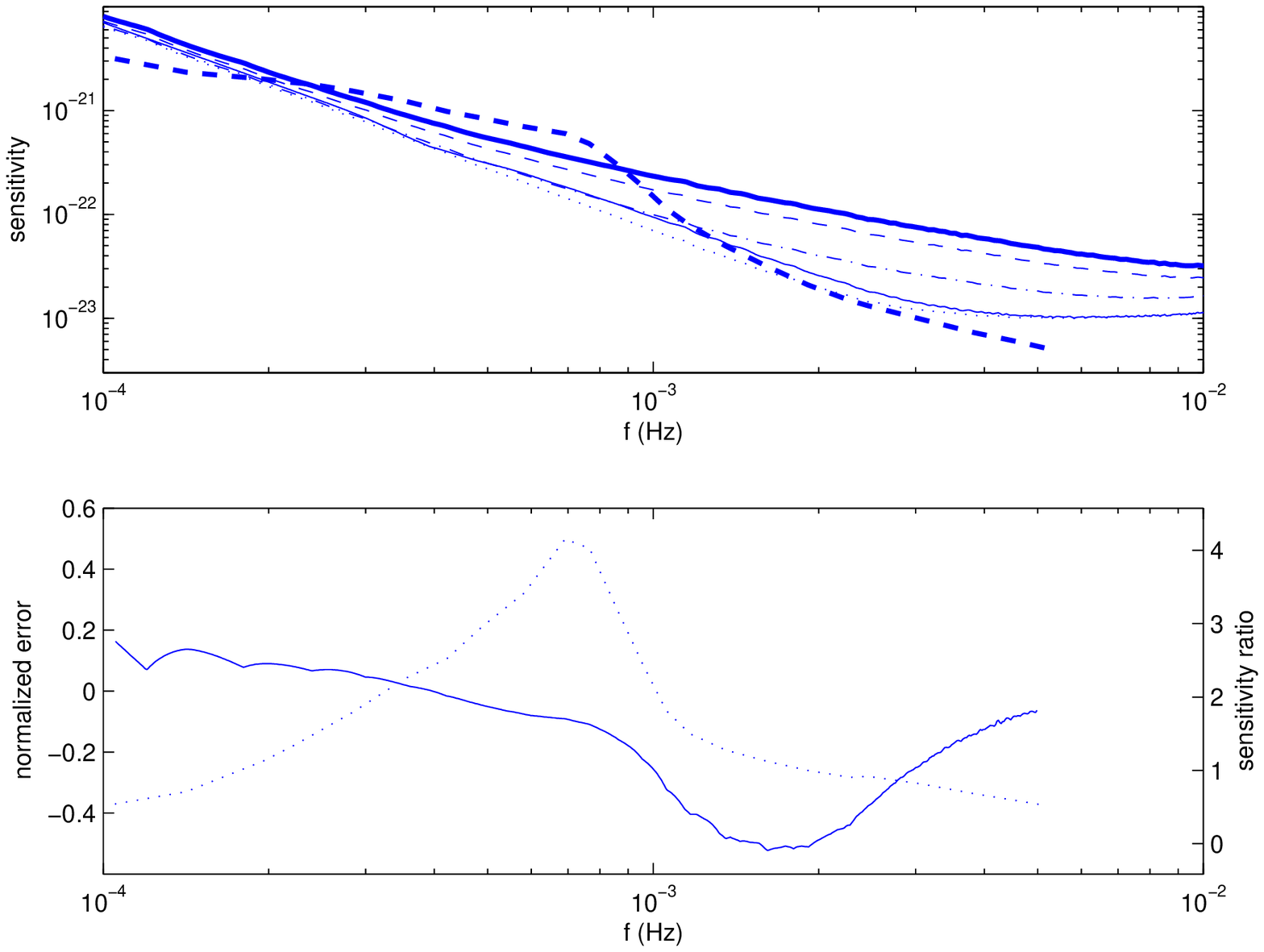}
\end{center}
\caption{{\it Top:} a comparison of our least square bound (thin continuous line) with the bound of \cite[Eq. (12)]{Tinto2000} (dash-dotted line for a lower bound on $V$ and $N$ ($S^0$ and $S^1$ in the notation of \cite{Tinto2000}) of 90\% of their design values, and the thin dashed line for a lower bound of 50\% of their design values), for the sensitivity of $S^{\rm ins}_{XX}$ (dotted line) with $\xi = \chi = 0$, for 1 year of observation and a signal to noise ratio of 5 (the sensitivity is averaged over sky position and polarization angle, as in \cite{TDIWP}). The thick continuous line shows the sensitivity for $16 S_{\zeta\zeta}$, the limit of the method of \cite{Tinto2000} when $V$ and $N$ are allowed to be vanishingly small. The thick dashed line shows the confusion noise from binaries calculated by \cite{BH1997} (dashed line in their Fig. 2). {\it Bottom:} the normalized systematic error on the sensitivity of $X$ (continuous line, left axis), computed from the least square bound minus the sensitivity of $S^{\rm ins}_{XX}$, divided by the strain in the stochastic background, and the ratio of the sensitivity of $S^{\rm ins}_{XX}$ to the strain in the stochastic background (dotted line, right axis).}
\label{fig:tintocomp}
\end{figure}

To give a specific example, we plot in Fig. \ref{fig:tintocomp} our
best-fitted spectrum and the estimated spectrum from \cite{Tinto2000}
for the $S^{\rm ins}_{XX}$ spectrum, assuming no correlations ($\chi =
\xi = 0$).  As shown in Eq. (\ref{eq:tintobound}), the spectrum
estimator from \cite{Tinto2000} requires a knowledge of how small the
acceleration and optical path noises can be.  Without reliable prior
information, the lower limit can only be assumed to be zero, and the
spectrum estimate is $\hat{S}^{\rm ins}_{XX} = 16 S_{\zeta\zeta}$,
which is considerably worse than the least-square spectrum.  One has to
be able to guarantee that the noise is at least larger than 90\% of
the design value so that the spectrum estimate performs as well as the
optimal least square estimator.

If the real stochastic background is as strong as the one plotted in
Fig. \ref{fig:tintocomp}, it will be detectable unambiguously by just 
comparing the measured spectrum $S_{XX}$ to sixteen times the measured
spectrum $S_{\zeta\zeta}$.  However, the measurement of the excess power at
each frequency of $S_{XX}$ will carry a systematic error from the
imperfect characterization of the noise.  In the model that we
considered above, this error would bring down the largest frequency at
which the background can be reliably estimated from $\sim 2$ mHz to
$\sim 1$ mHz.

To conclude, our bounds show that the spectra
relevant to LISA's data analysis can only be measured to relative errors of the order of 10\% if the estimation is restricted to using only the model-independent,
signal-independent quantities $S_{\theta\zeta}$, for $\theta = \alpha,
\beta, \gamma, \zeta$.
It should be possible to unambiguously detect
large stochastic backgrounds, by comparing $S_{XX}$ to $16
S_{\zeta\zeta}$, for instance.  Uncertainties in the characterization
of LISA's noise will principally introduce errors in the estimation of
the power spectrum of the stochastic background.  

\acknowledgments

This research was performed by the Jet Propulsion Laboratory,
California Institute of Technology, under contract with the National
Aeronautics and Space Administration.

\appendix
\section{SNR loss and degradation in parameter accuracy} \label{sladipa}
We present in this appendix a derivation of the formulae describing the degradation of the SNR and of the error on the estimated parameters of a GW signal due to errors in the estimation of the spectra and cross-spectra describing the noise in the LISA detector.
For matched filtering, the detection statistics is given by
\begin{equation}
\Lambda = \int df \vc{x} \vc{C}^{-1} \vc{\chi}^\dagger.
\end{equation}
The SNR $\rho$ for this statistics can easily be shown to be
\begin{equation}
\rho^2 = \frac{\left(\int df \vc{\chi} \vc{C}^{-1} \vc{\chi}^\dagger\right)^2}{\int df \vc{\chi} \vc{C}^{-1} \vc{C}_0 \vc{C}^{-1} \vc{\chi}^\dagger}, \label{eq:SNRapp}
\end{equation}
where the correlation matrix is assumed to contain an error as described by Eq. (\ref{eq:pertC}).
When the correlation matrix is perfectly known ($\vc{C}_1 = 0$), Eq.(\ref{eq:SNRapp}) reduces to Eq.(\ref{eq:bestsnr}).
The $\vc{C}_0$ matrix in the denominator of Eq.(\ref{eq:SNRapp}) comes from the expectation of $|\Lambda|^2$ when only the instrumental noise is present.

If $\vc{C}_1 \ll \vc{C}_0$, the Neumann expansion can be used to write
\begin{equation}
\vc{C}^{-1} = \vc{C}_0^{-1} - \vc{C}_0^{-1} \vc{C}_1 \vc{C}_0^{-1} + \vc{C}_0^{-1} \vc{C}_1 \vc{C}_0^{-1} \vc{C}_1 \vc{C}_0^{-1} + O(\vc{C}_1^3).
\end{equation}
Expanding Eq.(\ref{eq:SNRapp}) in the small matrix $\vc{C}_1$, straightforward but lengthy algebra shows that the first order terms vanish, and that
\begin{equation}
\rho^2 = \rho_0^2\left[
1 - \frac{1}{\rho_0^2} \int df \vc{\chi} \vc{C}_0^{-1} \vc{C}_1 \vc{C}_0^{-1} \vc{C}_1 \vc{C}_0^{-1} \vc{\chi}^\dagger
+ \frac{1}{\rho_0^4}\left(
\int df \vc{C}_0^{-1} \vc{C}_1 \vc{C}_0^{-1} \vc{\chi}^\dagger
\right)^2
\right], \label{eq:SNRreduction}
\end{equation}
where $\rho_0$ is the SNR achievable when $\vc{C}_1 = 0$.

If $\vc{C}$ is known up to a scale factor for all frequencies where the signal is present, for instance, then Eq.(\ref{eq:SNRreduction}) shows as expected that the SNR is unaffected.
If, however, the error on $\vc{C}$ is frequency dependent over the frequency range where the signal contributes significantly to the SNR, the filter used to construct $\Lambda$ will be far from optimal, and significant degradation in the SNR will occur.
Supposing for simplicity that the signal contributes uniformly to the SNR over a bandwidth $B$, and supposing that the error on the correlation matrix is $\vc{C}_1 = \epsilon \vc{C}_0$ over a bandwidth $b$ included in the signal frequency range, then
\begin{equation}
\rho^2 \simeq 
\rho_0^2\left[ 
1 - \epsilon^2 \left(\frac{b}{B} - \frac{b^2}{B^2} \right)
\right].
\end{equation}
The SNR losses will be the smallest when the error on $\vc{C}$ is uniform over most of the bandwidth of the signal, as it is the case for narrow-band signals, and obviously when these errors are small where the signal is large compared to the instrumental noise.
The worst case scenario is a signal with $b/B=1/2$, in which case the fractional SNR loss is
\begin{equation}
\sqrt{\frac{\rho_0^2 - \rho^2}{\rho_0^2}} \simeq \frac{\epsilon}{2}. \label{eq:wcsnr}
\end{equation}

As similar perturbation analysis can be performed for the parameter estimation problem, by expanding Eqs.(\ref{eq:optestimator}) and (\ref{eq:fisherc}) to first order in $\vc{C}_1$, and then computing $\cov \hat{\vc{\lambda}}$ as defined in Eq.(\ref{eq:crbound}).
The inverse of the perturbed Fisher matrix given in Eq.(\ref{eq:fisherc}) is given by
\begin{equation}
\vc{I}^{-1} = \vc{I}_0^{-1} + \vc{I}_0^{-1} \frac{\partial\tilde{\vc{\chi}}}{\partial\vc{\lambda}} \vc{C}_0^{-1} \vc{C}_1 \vc{C}_0^{-1} \frac{\partial\tilde{\vc{\chi}}^\dagger}{\partial\vc{\lambda}} \vc{I}_0^{-1} + O(\vc{C}_1^2),
\end{equation}
where $\vc{I}_0^{-1}$ is the achievable covariance when the
correlation matrix is known exactly ($\vc{C}_1 = 0$).
Using this and the first order expansion of the inverse of the correlation matrix into Eq.(\ref{eq:optestimator}) gives four terms that are first order in $\vc{C}_1$.
These four terms can be simplified using the following identities
\begin{eqnarray}
\vc{I}_0^{-1} = \vc{I}_0^{-1} E\left[
{\rm Re}\left[\frac{\partial\tilde{\vc{\chi}}}{\partial\vc{\lambda}} \vc{C}^{-1} (\tilde{\vc{x}} - \tilde{\vc{\chi}})^\dagger \right]
{\rm Re}\left[\frac{\partial\tilde{\vc{\chi}}}{\partial\vc{\lambda}} \vc{C}^{-1} (\tilde{\vc{x}} - \tilde{\vc{\chi}})^\dagger \right]^T
\right] \vc{I}_0^{-1} \\
E[(\vc{x}-\vc{\chi})^\dagger (\vc{x}-\vc{\chi})] = \vc{C}_0 \\
E[(\vc{x}-\vc{\chi})^T (\vc{x}-\vc{\chi})] = 0.
\end{eqnarray}
The result is that the terms that are first order in $\vc{C}_1$ all cancel, as they did for the SNR.
The increase in the covariance of the estimated quantities is therefore at most second order in the errors on the measurement of the correlation matrix.


\begin{references}

\bibitem{galbin} C. R. Evans, Jr. I. Iben, and L. Smarr, Astrophys. J. {\bf 323}, 129 (1987) ;  D. Hils, P. L. Bender, and R. F. Webbink, Astrophys. J. {\bf 360}, 75 (1990) ; G. Nelemans, L. R. Yungelson, and S. F. Portegies Zwart, Astron. Astrophys. {\bf 375}, 890 (2001).

\bibitem{LSR} LISA Science Requirements document (2002).

\bibitem{TDI} M. Tinto and J. W. Armstrong, Phys. Rev. D {\bf 59}, 102003 (1999) ; F. B. Estabrook, M. Tinto, and J. W. Armstrong, Phys. Rev. D {\bf 62}, 042002 (2000) ; M. Tinto, F. B. Estabrook, and J. W. Armstrong, Phys. Rev. D {\bf 65}, 082003 (2002) ; M. Tinto, D. A. Shaddock, J. Sylvestre, and J. W. Armstrong, Phys. Rev. D {\bf 67}, 122003 (2003).

\bibitem{aet99} J. W. Armstrong, F. B. Estabrook and M. Tinto, Astrophys. J. {\bf 527}, 814 (1999).

\bibitem{TDIWP} M. Tinto, F. B. Estabrook and J. W. Armstrong, ``Time-Delay Interferometry and LISA's Sensitivity to Sinusoidal Gravitational Waves'', technical document of the LISA project (2002). Available at http://www.srl.caltech.edu/lisa/tdi\_wp/LISA\_Whitepaper.pdf .

\bibitem{Shaddock03} D.A. Shaddock, gr-qc/0306125 (2003).

\bibitem{CH} N. Cornish and R.W. Hellings, gr-qc/0306096 (2003).

\bibitem{STEA} D.A. Shaddock, M. Tinto, F.B. Estabrook, and
  J.W. Armstrong, gr-qc/0307080 (2003).

\bibitem{pmla02} T. A. Prince, M. Tinto, S. L. Larson, and J. W. Armstrong, Phys. Rev. D {\bf 66}, 122002 (2002).

\bibitem{tk03} A. Kr\'olak and M. Tinto, gr-qc/0302013.

\bibitem{Tinto2000} M. Tinto, J. W. Armstrong, and F. B. Estabrook, Phys. Rev. D {\bf 63}, 021101R (2001).

\bibitem{Hogan2001} C. J. Hogan and P. L. Bender, Phys. Rev. D {\bf 64}, 062002 (2001).

\bibitem{PPA} LISA System and Technology Study Report (2000).

\bibitem{bender} P. L. Bender, ``Suggested Candidate for LISA Error Allocation Budget'', unpublished document (2003).

\bibitem{xcpap} J. Sylvestre and M. Tinto, in preparation.

\bibitem{NR} W. H. Press et al., {\it Numerical recipes in C: the art of scientific computing}, Cambridge University Press (1988).

\bibitem{BH1997} P. L. Bender, D. Hils, Class. Quantum Grav. {\bf 14}, 1439 (1997).

\end{references}
\end{document}